\documentclass{ws-mpla-hep}
 \usepackage{amsmath,amssymb}

\def\muF{\relax\ifmmode\mu_\text{F}^2\else{$\mu_\text{F}^2${ }}\fi}
\def\muR{\relax\ifmmode\mu_\text{R}^2\else{$\mu_\text{R}^2${ }}\fi}

\begin{document}

\def\preprint{RUB-TPII-13/09}

\markboth{A.~P.~Bakulev, A.~V.~Pimikov, and N.~G.~Stefanis}
{Pion FF in NLC QCD SRs and in LD}

\catchline{}{}{}{}{}

\title{Pion Form Factor in QCD Sum Rules with Nonlocal Condensates
       and in the Local-Duality Approach}

\author{\footnotesize A.~P.~BAKULEV\footnote{
 Talk presented at Workshop
 ``Recent Advances in Perturbative QCD and Hadronic Physics'',
 20--25 July 2009, ECT*, Trento (Italy),
 in Honor of Prof. Anatoly Efremov's 75th Birthday Celebration.}}

\address{Bogoliubov Laboratory of Theoretical Physics,
         JINR,
         Dubna, 141980, Russia\\
         bakulev@theor.jinr.ru}

\author{\footnotesize A.~V.~PIMIKOV}

\address{Bogoliubov Laboratory of Theoretical Physics,
         JINR,
         Dubna, 141980, Russia\\
         pimikov@theor.jinr.ru}

\author{\footnotesize N.~G.~STEFANIS\footnote{
         Also at Bogoliubov Laboratory of Theoretical Physics,
         JINR,
         Dubna, 141980, Russia}}

\address{Institut f\"{u}r Theoretische Physik II,
         Ruhr-Universit\"{a}t Bochum, D-44780 Bochum, Germany\\
         N.G.Stefanis@tp2.ruhr-uni-bochum.de}

\maketitle

\pub{Received (Day Month Year)}{Revised (Day Month Year)}

\begin{abstract}
We discuss the QCD sum-rule approach
for the spacelike electromagnetic pion form factor
in the $O(\alpha_s)$ approximation.
We show that the nonlocality of the condensates is a key point
to include nonperturbative contributions to the pion form factor.
We compare our results with the Local-Duality predictions
and show that the continuum threshold $s_0(Q^2)$ parameter is highly
underestimated in the Local-Duality approach at $Q^2\gtrsim 2$~GeV$^2$.
Using our fit for this parameter,  $s_0^\text{LD}(Q^2)$,
and applying the fractional analytic perturbation theory,
we estimate with an accuracy of the order of 1\%
the $O(\alpha_s^2)$ contribution to the pion's form factor.
\end{abstract}

\ccode{PACS Nos.: 12.38.Aw, 12.38.Bx, 13.40.Gp}

\section{A tribute to Prof.~Efremov's 75\lowercase{th} birthday}
We are happy to point out in this Festschrift,
the influence of Prof. Efremov's work on our own research activities.

A.~V.~Efremov is one of the inventors of
the factorization theorems
in quantum field theory,
which form the basis of perturbative QCD applications
in exclusive\cite{ER80}
and inclusive\cite{ER80tmf1} reactions
with hadrons.
Without these tools,
the experimental verification of QCD would be impossible.
In cooperation with his then student A.~V.~Radyushkin
he generalized the factorization theorems
for the meson form factors,
linking diagrammatic techniques with the operator product expansion (OPE).
And all these achievements were based on previous investigations
by Efremov.\cite{Efr74yad,EG74fph}

Moreover, Efremov and Radyushkin
have diagonalized the anomalous-dimensions matrix for meson operators
(in leading order) in terms of Gegenbauer polynomials
and first obtained the asymptotic distribution amplitude (DA)
$\varphi(x,\mu^2\to\infty)\to\varphi^{as}(x)=6x(1-x)$.\cite{ER80,LB79}
Factorization theorems\cite{ER80,CZ77,LB79} make it possible to
calculate various hard processes in QCD involving mesons,
in which the meson DAs enter as the central nonperturbative input.
The title of the paper\cite{ER80}
``\textit{Factorization and asymptotic behaviour of pion form factor in QCD}''
shows explicitly that the main interest
was focused on the pion form factor (FF)
for which the leading asymptotics at large $Q^2$ was revealed:
$F_\pi(Q^2) \to 8\,\pi\,f_\pi^2\,\alpha_s(Q^2)/Q^2$.
The precise value of $Q^2$ at which this asymptotic regime starts
to prevail cannot be determined accurately:
the estimates range from $100$~GeV$^2$ in Refs.\ \refcite{IL84,JK93,BLM07}
down to values around
$20$~GeV$^2$ in Refs.\ \refcite{SSK,BPSS04}.
However, all these values are still rather far away
from the capabilities of any operating or planned accelerator facility.

To estimate the pion FF at intermediate $Q^2$,
one actually needs to employ some nonperturbative approach,
like the QCD Sum Rules\cite{NR82,IS82} or Local Duality.\cite{NR83,BRS00,BPSS04,BLM07}
We present here our recent results
on the pion FF
obtained with QCD Sum Rules (SRs) with nonlocal condensates (NLCs).

\section{Pion form factor in the QCD SRs approach}
The spacelike electromagnetic pion form factor (FF) describes the
scattering of charged particles off the pion by exchanging a photon
and is defined by the following matrix element:
\begin{eqnarray*}
  \langle{\pi^{+}(P^{\prime})| J_{\mu}(0) | \pi^{+}(P)\rangle}
    =  {\left( P + P^{\prime}\right)}_{\mu} F_{\pi}(Q^{2})\,.
\end{eqnarray*}
Here $J_\mu$ is the electromagnetic current
and  $q$ is the photon momentum  $q^2=(P^{\prime} - P)^2=-Q^2<0$
in the spacelike region.
To extract information about the pion form factor in the QCD SR approach,
one needs to investigate
the Axial-Axial-Vector (AAV) correlator of
the EM current
$J^{\mu}(x)=e_u\,\overline{u}(x)\gamma^\mu u(x)
           +e_d\,\overline{d}(x)\gamma^\mu d(x)$
(here $e_u=2/3$ and $e_d=-1/3$ stand
 for the electric charges of the $u$ and the $d$ quarks)
and two axial-vector currents $J_{5\alpha}(x)=   \overline{d}(x)\gamma_5\gamma_\alpha u(x)$:
\begin{eqnarray*}
    \int\!\!\!\int\!\!d^4x\,d^4y\,e^{i(qx-P^\prime y)}
    \langle{0|T\!\!\left[J^{+}_{5\beta}(y) J^{\mu}(x) J_{5\alpha}(0)\right]\!\!|0
           }\rangle\,.
\end{eqnarray*}

Using the standard QCD SRs technique\cite{NR82,IS82}
in conjunction with the concept of nonlocal condensates,\cite{MR-NLC,BR91,BMS01,BP06}
we obtain the following SR:
\begin{eqnarray}
\label{eq:ffQCDSR}
  f_{\pi}^2\,F_{\pi}(Q^2)
  &=& \int\limits_{0}^{s_0}\!\!\!\int\limits_{0}^{s_0}\!ds_1\,ds_2\
           \rho_3(s_1, s_2, Q^2)\,
            e^{-(s_1+s_2)/M^2}\nonumber\\
  &+& \Phi_\text{G}(Q^2,M^2)
    + \Phi_{\langle\bar{q}q\rangle}(Q^2,M^2)\,,
\end{eqnarray}
where $M^2$ is the Borel parameter,
the term $\Phi_\text{G}(Q^2,M^2)$ represents
the gluon-condensate contribution,
while the quark-condensate contribution
$\Phi_{\langle\bar{q}q\rangle}(Q^2,M^2)$
consists of the sum of the four-quark condensate $\Phi_\text{4Q}(Q^2,M^2)$,
the bilocal vector-quark condensate $\Phi_\text{2V}(Q^2,M^2)$,
and the antiquark-gluon-quark condensate $\Phi_{\bar qAq}(Q^2,M^2)$.

The three-point spectral density is of the form
\begin{eqnarray}
 \label{eq:SpDen.pert}
  \rho^{(1)}_3(s_1, s_2, Q^2)
   = \left[\rho_3^{(0)}(s_1, s_2, Q^2)
        + \frac{\alpha_s(Q^2)}{4\pi}\,
           \Delta\rho_3^{(1)}(s_1, s_2, Q^2)
    \right]\,,
\end{eqnarray}
where the leading-order spectral density
has been calculated long ago,\cite{NR82,IS82}
while the next-to-leading order (NLO) version
$\Delta\rho_3^{(1)} (s_1, s_2, Q^2)$
has been derived recently in Ref.\ \refcite{BO04}.
The contribution from higher resonances is usually taken into account
in the form of
$\rho_\text{HR}(s_1, s_2) =
 \left[1-\theta(s_1<s_0)\theta(s_2<s_0)\right]\,
 \rho_3(s_1, s_2, Q^2)$
and contains the continuum threshold parameter $s_0$.
We use in the perturbative spectral density
the analytic version of the running coupling
that avoids Landau singularities by construction
(see for reviews in  Refs.\ \refcite{SS97-06,AB08,Ste09}):
\begin{eqnarray}
 \label{eq:alphaS}
  \alpha_s(Q^2)
   &=&\frac{4\pi}{b_0}
       \left(\frac{1}{\ln(Q^2/\Lambda_{\text{QCD}}^2)}
           - \frac{\Lambda_{\text{QCD}}^2}{Q^2-\Lambda_{\text{QCD}}^2}
                \right)
\end{eqnarray}
with $b_0=9$ and $\Lambda_\text{QCD}=300$~MeV.

\section{Nonlocal condensates in QCD SRs for the pion FF}
In perturbation theory,
the vacuum coincides with the ground state
of the free-field theory;
hence the expectation value of the normal product is zero.
Therefore,
there are no condensate terms in perturbation theory.
However, in the physical vacuum this is not the case.
For this reason, in the standard QCD SR approach
the nonzero quark condensate $\langle{\bar q q\rangle}\equiv\langle{\bar q_A(0) q_A(0)\rangle}$
appears.
The value of this constant was defined through comparison with experimental data
for the $J/\psi$-meson.\cite{SVZ}
Assuming a small coordinate dependence,
the quark condensate can be represented by the first two terms
of the Taylor expansion
\begin{eqnarray}
\label{eq:loc.cond}
 \langle{\bar q_B(0)\,q_A(x)\rangle}
  = \frac{\delta_{AB}}{4}\,
     \bigg[ \langle{\bar q q\rangle}
          + \ldots
     \bigg]
  + i\,\frac{\widehat{x}_{AB}}{4}\,
       \frac{x^2}{4}\,
        \bigg[\frac{2\alpha_s\pi\langle{\bar qq}\rangle^2}{81} + \ldots
        \bigg]\,,
\end{eqnarray}
where we kept the scalar and vector parts apart.
Note that the condensates in this representation are local.

As has been shown in Refs.\ \refcite{MR-NLC,BR91,MS93,BPS09},
the local approximation (\ref{eq:loc.cond}) is not reasonable
for studying FFs and DAs.
The reason is the unphysical behavior of the local condensate (\ref{eq:loc.cond})
at large $x^2$,
which entails a constant scalar term and a vector term
that is even growing with the distance between the quarks $x^2$.
As a result, the nonperturbative part of the OPE linearly increases
with the momentum $Q^2$:
$\left(c_1 + Q^2/M^2\right)$,
where $c_1$ is a dimensionless constant (not depending on $Q^2$).
At the same time, the perturbative part decreases with $Q^2$,
hence generating an inconsistency of the SR
at intermediate and large $Q^2$.
Therefore, we can not rely upon the obtained SR for the pion FF
for momentum values $Q^2>3$~GeV$^2$.

In order to improve the $Q^2$ dependence,
one needs to modify the model
of the quark-condensate behavior
at large distances.
Indeed, lattice simulations\cite{DDM99,BM02}
and instanton models\cite{DEM97,PW96}
indicate a decrease of the scalar quark condensate
with increasing interquark distance,
thus confirming the approach of NLCs.\cite{MR-NLC}
The main strategy of the NLC SRs\cite{MR-NLC,BR91,MS93}
is to avoid the original Taylor expansion and
to deal directly with the NLCs by introducing model functions
that describe the coordinate dependence of the condensates.

In the NLC approach the bilocal quark-antiquark condensate
has the following form:\footnote{%
We use the Euclidean interval $x^2 = -x_0^2-\vec{x}^2<0$.
As usual in the QCD SR approach,
the Fock--Schwinger gauge is used.
For this reason, all string connectors
${\mathcal C}(x,0) \equiv
 {\mathcal P}
  \exp\!\left[-ig_s\!\!\int_0^x t^{a} A_\mu^{a}(y)dy^\mu\right]=1
$.}
\begin{eqnarray}
 \langle{\bar{q}_A(0) q_B(x)}\rangle
  = \frac{1}{4}
     \int_0^\infty\!\!
     \Big[\delta_{BA}\,\langle{\bar{ q}q}\rangle\,f_S(\alpha)
         - iA_0\,\widehat{x}_{BA}\,f_V(\alpha)
     \Big]\,
      e^{\alpha x^2/4}d\alpha\,,
\end{eqnarray}
which, for the most general case,
is parameterized by the distribution functions
$f_S(\alpha)$ and $f_V(\alpha)$,
with $A_0=2\alpha_s\pi\langle{\bar{q}q}\rangle^2/81$.
The explicit form of these functions must be taken
from a concrete model
of the nonperturbative QCD vacuum.
In the absence of an exact QCD solution,
it was proposed\cite{MR-NLC}
to use the first nontrivial approximation
which takes into account only the finite width of the spatial
distribution of the vacuum quarks:
$f_S(\alpha) = \delta\left(\alpha-\lambda_q^2/2\right)$.
This generates a Gaussian form
of the NLC in the coordinate representation:
$\langle{\bar{q}_A(0) q_A(x)}\rangle = \langle{\bar{q}q}\rangle e^{-|x|^2\lambda_q^2/8}$,
which leads to the following form of the condensate contributions to the FF:
$\left(c_1 + Q^2/M^2\right)\,e^{- c_2 Q^2\lambda_q^2/M^4}$,
where $c_i$ are dimensionless constants not depending on $Q^2$.
Thus, the nonlocality of the vacuum condensates
generates a decreasing behavior of the nonperturbative part
of the FF at large $Q^2$.

The same technique is applied
in the case of the mixed quark-gluon condensate,
$\langle{\bar{q}_B(0)(-g A^a_\nu(y)\,t^a)q_A(x)}\rangle$.
There are two models for this condensate:
the minimal and the improved one,
see for details in Ref.\ \refcite{BP06,BPS09}.
The nonlocal gluon-condensate contribution produces
a very complicated expression.
But owing to its smallness,
we can model the nonlocality of the gluon-condensate
in analogy to the quark case,
using an exponential factor,\cite{BR91,MS93}
notably, $e^{-\lambda_g^2 Q^2/M^4}$.

\begin{figure}[h]
 \centerline{\psfig{file=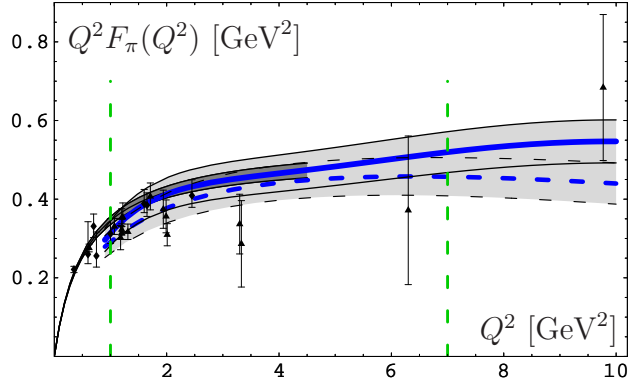,width=0.65\textwidth}}
  \caption{Scaled pion form factor $Q^2 F_{\pi}(Q^2)$ for the minimal NLC model
   (shown as a thick broken line inside the shaded band delimited by the dashed
    lines which denote the uncertainty range). The improved NLC model is represented
   by a solid line inside the shaded band within the solid lines
   ($\lambda_q^2=0.4$~GeV$^2$). The two broken vertical lines mark the region,
   where the influence of the particular Gaussian model used to parameterize
   the QCD vacuum structure in the NLC QCD SRs is not so strong. The recent lattice
   result of Ref.\ \protect\refcite{Brommel06} is shown as a monopole fit with error bars
   between the two thick lines at lower $Q^2$.\protect\label{fig:Q2FF}}
\end{figure}

The described NLC QCD SR approach was used
for the calculation of the pion FF
in Ref.\ \refcite{BPS09}.
This method yields predictions
for the spacelike pion form factor (see Fig.~\ref{fig:Q2FF})
that compare well with the experimental data
of the Cornell\cite{FFPI-Cornell} (triangles)
and the JLab Collaborations\cite{JLab08II} (diamonds)
in the momentum region currently accessible to experiment.
These predictions cover also the range of momenta
to be probed by the $12$~GeV$^2$ upgraded CEBAF accelerator at the Jefferson Lab
in the near future.
This planned high-precision measurement of the pion FF at JLab
will certainly help to check the quality of the discussed NLC models.

\section{Pion FF in the Local-Duality approach}
 \label{sec:PiFF.LD}
The LD SR\cite{NR82,Rad95} is constructed
from the original QCD SR
in the $M^2\rightarrow \infty$ limit.
For this reason it has no condensate contributions.
The main nonperturbative ingredient in this approach
is the effective continuum threshold $s_0^\text{LD}$ --- it inherits
all the nonperturbative information from the original QCD SR.
At the $(l+1)$-loop order we have
\begin{eqnarray}
 \label{eq:FF.LD}
  F_{\pi}^{\text{LD};(l)}(Q^2,S)
   \equiv
    \int\limits_{0}^{S}\!\!\!\int\limits_{0}^{S}\!\!
     \rho_3^{(l)}(s_1, s_2, Q^2)\,
      \frac{ds_1\,ds_2}{f_{\pi}^2}\,,
\end{eqnarray}
where $S$ should be substituted by the LD effective threshold,
$s_0^{\text{LD};(l)}(Q^2)$,
and $\rho_{3}^{(l)}(s_1, s_2, Q^2)$
is the three-point $(l+1)$-loop spectral density.
In leading order the integration can be done analytically
and yields
\begin{eqnarray*}
 F_{\pi}^{\text{LD};(0)}(Q^2,S)
  = \frac{S}{4\pi^2f_\pi^2}\,
    \left(1 - \frac{Q^2+6S}{Q^2+4S}\,\sqrt{\frac{Q^2}{Q^2 + 4 S}}\,
    \right)\,.
\end{eqnarray*}
The LD prescription for the corresponding correlator\cite{SVZ,Rad95}
implies the relations
\begin{subequations}
\label{eq:LD.s0}
\begin{eqnarray}
\label{eq:LD.s0.LO}
 s_0^{\text{LD};(0)}(0)
  = 4\,\pi^2\,f_{\pi}^2 \simeq 0.7~\text{GeV}^2
\end{eqnarray}
and
\begin{eqnarray}
\label{eq:LD.s0.NLO}
 s_0^{\text{LD};(1)}(0)
  = \frac{4\,\pi^2\,f_{\pi}^2}
         {1+\alpha_s(Q_{0}^2)/\pi}
  \simeq 0.6~\text{GeV}^2\,,~~~
\end{eqnarray}
\end{subequations}
where $Q_{0}^2$ is of the order of $s_0^{\text{LD};(0)}(0)$.
This prescription is a strict consequence of the Ward identity
for the AAV correlator due to the vector-current conservation.
In principle, the $Q^2$ dependence of the LD parameter
$s_0^{\text{LD}}(Q^2)$ (\ref{eq:FF.LD})
should be determined from the QCD SR at $Q^2\gtrsim1$~GeV$^2$.
But as explained in Refs.\ \refcite{IS83,NR84,BPS09},
the standard QCD SR becomes unstable at $Q^2>3$~GeV$^2$
because of the appearance
of terms in the condensate contributions linearly growing
with $Q^2$.
For this reason, this dependence was known only for
$Q^2\leq 3$~GeV$^2$ and, therefore, most authors usually used the
constant approximation
$s_0^{\text{LD};(0)}(Q^2)\simeq s_0^{\text{LD};(0)}(0)$,
like in Refs.\ \refcite{NR82,BRS00,BPSS04,BO04},
or a slightly $Q^2$-dependent
approximation
$ s_0^{\text{LD};(1)}(Q^2)
\simeq
 4\,\pi^2\,f_{\pi}^2/(1+\alpha_s(Q^2)/\pi)
$,
like in Ref.\ \refcite{BLM07}.

But now, due to the knowledge of the NLC QCD SR prediction\cite{BPS09}
for the pion FF for $Q^2=1-10$~GeV$^2$,
we can estimate the effective LD thresholds
$s_{0}^\text{LD}(Q^2)$,
which reproduce these predictions
in the LD approach,
for the two used Gaussian models of the QCD vacuum,
the minimal and the improved one.
Results are shown in Fig.~1.
They can be represented in this $Q^2$ range
by the following interpolation formulas:
\begin{subequations}
\begin{eqnarray}
 \label{eq:FF.LD.s0.App}
  s_{0,\text{min}}^\text{LD}(Q^2=x~\text{GeV}^2)
  &=& 0.57+0.307\,\tanh(0.165\,x)
          -0.0323\,\tanh(775\,x)\,;\\
  s_{0,\text{imp}}^\text{LD}(Q^2=x~\text{GeV}^2)
  &=& 0.57+0.461\,\tanh(0.0954\,x)\,.~~~
\end{eqnarray}
\end{subequations}
We see that $s_{0}^\text{LD}(Q^2)$
in the mentioned range of $Q^2$
is a monotonically increasing function.
Therefore, $s_{0}^\text{LD}(Q^2)\neq s_{0}^\text{SR}(Q^2)\approx0.7$~GeV$^2$
and,
due to this difference,
the LD approaches of Refs.\ \refcite{BPSS04,BO04,BLM07}
produce significantly lower predictions for $Q^2\,F_\pi(Q^2)$
as compared with QCD SRs with NLCs.

\begin{figure}[htb]
 \centerline{\psfig{file=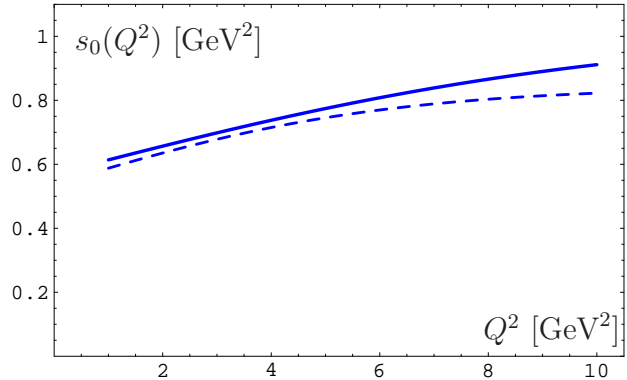,width=0.65\textwidth}}
   \caption{Effective continuum thresholds
     $s_{0,\text{imp}}^\text{LD}(Q^2)$ (solid line) and
     $s_{0,\text{min}}^\text{LD}(Q^2)$ (dashed line) that approximate
     the NLC QCD SR results using the LD $O(\alpha_s(Q^2))$-formulas.
    \protect\label{fig:LD.s0}}
\end{figure}

\section{Using Fractional Analytic Perturbation Theory for two-loop estimates}
 \label{sec:FAPT}
To estimate the Next-to-Next-to-Leading-Order (NNLO),
i.e., the two-loop,
contribution
to the pion FF in the QCD SR approach,
one needs to calculate the three-loop spectral density
$\rho_{3}^{(3)}(s_1, s_2, Q^2)$ --- a complicated task.
We want to avoid this calculation
and suggest instead to use the known collinear two-loop result,
the LD model for the soft part with an improved parameter $s_0^\text{LD}(Q^2)$,
and the matching procedure of Ref. \refcite{BPSS04}.
We also apply Fractional Analytic Perturbation Theory (FAPT)
for the two-loop collinear expression
in order to have an approximate independence
with respect to the renormalization and factorization scales,
see in Refs.\ \refcite{BPSS04,AB08}.

To combine the dominant, at small $Q^2\leq 1$~GeV$^2$, LD model
for the soft part,
$F_{\pi}^{\text{LD},(0)}(Q^{2})$,
with the perturbative hard-rescattering part,
$F_{\pi}^{\text{pQCD},(2)}(Q^2)$
(which provides the leading perturbative
 $O(\alpha_s)+O(\alpha_s^2)$ corrections
 and is dominant at large $Q^2\gg1$~GeV$^2$),
in such a way as to ensure the validity of the Ward identity (WI)
$F_{\pi}^{\text{WI};(2)}(0)=1$,
we apply the matching procedure,
introduced in Ref.\ \refcite{BPSS04}:
\begin{eqnarray}
  F_{\pi}^{\text{WI};(2)}(Q^{2})
  = F_{\pi}^{\text{LD},(0)}(Q^{2})
  + \left(\frac{Q^2}{2s_0^{(2)}+Q^2}\right)^2
     F_{\pi}^{\text{pQCD},(2)}(Q^2)
 \label{eq:Fpi-Mod.NNLO}
\end{eqnarray}
with $s_0^{(2)}\simeq0.6$~GeV$^2$.
To test the quality of the matching prescription
given by Eq.\ (\ref{eq:Fpi-Mod.NNLO}),
we compare it with
the LD model (\ref{eq:FF.LD})
evaluated at the $O(\alpha_s)$-approximation.\cite{BO04,BLM07}
To this end, we construct the analogous $O(\alpha_s)$-model
$F_{\pi}^{\text{WI};(1)}(Q^{2})$,
where we substitute $F_{\pi}^{\text{pQCD},(2)}(Q^2)$
by $F_{\pi}^{\text{pQCD},(1)}(Q^2)=2\,\alpha_s(Q^2)\,s_0^{\text{LD};(0)}(0)/\pi\,Q^2$
and employ the same prescription for the effective LD threshold
as in Refs.\ \refcite{BLM07},
i.e., Eq.\ (\ref{eq:LD.s0.NLO}).
The key feature of this matching recipe is that
it uses the information on $F_{\pi}(Q^2)$
in the two asymptotic regions:
\begin{enumerate}
 \item $Q^2\to0$,
   where the Ward identity dictates $F_{\pi}(0)=1$
   and,
   hence, $F_{\pi}(Q^2)\simeq F_{\pi}^{\text{LD},(0)}(Q^2)$,
 \item $Q^2\to\infty$,
   where $F_{\pi}(Q^2)\simeq F_{\pi}^{\text{pQCD},(1)}(Q^2)$
\end{enumerate}
in order to combine properly
the hard tail of the pion FF with its soft part.
Numerical analysis shows
that the applied prescription yields
a pretty accurate result,
with a relative error
varying in the range 5\% at $Q^2=1$~GeV$^2$ to 9\% at
$Q^2=3-30$~GeV$^2$.
Note here that this recipe was proposed
without the knowledge
of the exact two-loop spectral density ---
this appeared later.\cite{BO04}

Now, knowing the exact expression for the spectral density $\rho_{3}^{(1)}(s_1, s_2, Q^2)$,
we can improve
the representation of the LD part
by taking into account
the leading $O(\alpha_s)$ correction in the electromagnetic vertex.
To achieve this goal,
we suggest the following improved WI model:
\begin{eqnarray}
  F_{\pi;\text{imp}}^{\text{WI};(1)}(Q^{2},S)
   = F_{\pi}^{\text{LD};(0)}(Q^{2},S)
   &\!+\!& \frac{S}{4\pi^2f_\pi^2}\,
      \frac{\alpha_s(Q^2)}{\pi}\,
       \left(\frac{2S}{2S+Q^2}\right)^2
    \nonumber\\
   &\!+\!& \frac{S}{4\pi^2f_\pi^2}\,
       F^{\text{pQCD},(1)}_\pi(Q^2)\,
       \left(\frac{Q^2}{2S+Q^2}\right)^2~~~
 \label{eq:Fpi-Mod.imp}
\end{eqnarray}
with the subsequent substitution $S\to s_0^{\text{LD};(1)}(Q^2)$.
Numerical evaluation of this new WI model in comparison with the
exact LD result in the one-loop approximation shows
that the quality of the matching condition is improved:
the relative error is reduced,
reaching only 4\% at $Q^2=1-10$~GeV$^2$.

We construct the two-loop WI model
$F_{\pi}^{\text{WI};(2)}(Q^{2},s_0^{\text{LD};(2)}(Q^2))$
for the pion FF to obtain
\begin{eqnarray}
  F_{\pi}^{\text{WI};(2)}(Q^{2},S)
  = F_{\pi}^{\text{LD};(0)}(Q^{2},S)
  &\!+\!& \frac{S}{4\pi^2f_\pi^2}\,
           \frac{\alpha_s(Q^2)}{\pi}\,
            \left(\frac{2S}{2S+Q^2}\right)^2
    \nonumber\\
  &\!+\!& \frac{S}{4\pi^2f_\pi^2}\,
           F^{\text{FAPT},(2)}_\pi(Q^2)\,
           \left(\frac{Q^2}{2S+Q^2}\right)^2\,,~~~
 \label{eq:Fpi.WI-Mod.NNLO}
\end{eqnarray}
where $F^{\text{FAPT},(2)}_\pi(Q^2)$ is the analyticized
expression generated from $F^{\text{pQCD},(2)}_\pi(Q^2)$
using FAPT (see Refs.\ \refcite{BMS-APT,BKS05,AB08})
to get a result which appears to be very close
to the outcome of the default scale setting
($\muR=\muF=Q^2$),
investigated in detail in Ref.\ \refcite{BPSS04} in the APT approach.
FAPT is needed here in order to obtain analytic expressions
for the pion FF,
using two possible options for the factorization scale:
\begin{enumerate}
 \item[(i)] For $\muF=Q^2$,
    there appear factors of the type
    $\left[\alpha_s(Q^2)\right]^{\nu}$
    with fractional powers $\nu=\gamma_n/(2\,b_0)$
    due to the evolution of the pion distribution amplitude;
 \item[(ii)] For $\muF=\textsl{const.}$,
    the factor
    $\left[\alpha_s(Q^2)\right]^{2}\ln(Q^2/\muF)$ appears.
\end{enumerate}
In any case, the NNLO correction
involves the analytic image of the second power of the coupling,
$\mathcal A_2(Q^2)$.
For this reason, we call
the whole $F^{\text{FAPT},(2)}_\pi(Q^2)$ term
the $O(\mathcal A_2)$ contribution.\vspace*{-3mm}

\begin{figure}[h]
 \centerline{\psfig{file=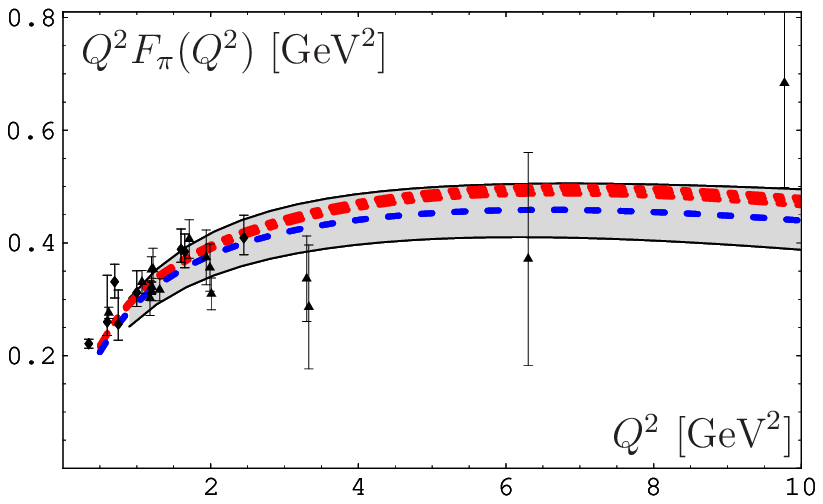,width=0.47\textwidth}
          ~~~\psfig{file=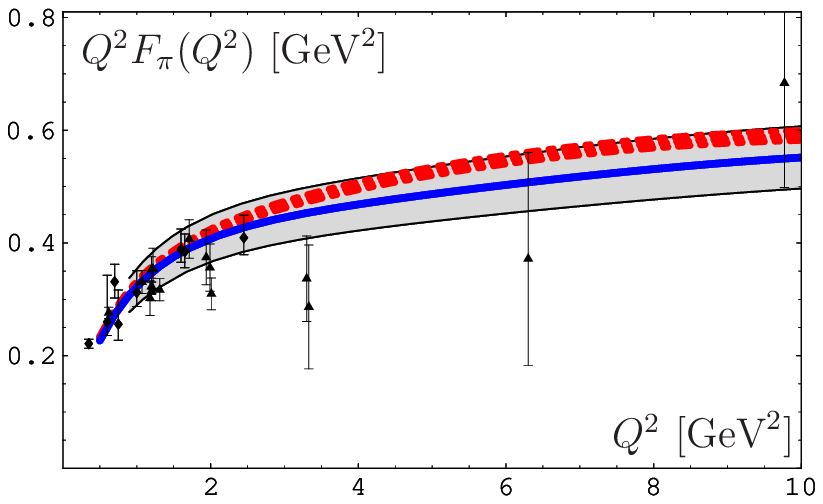,width=0.47\textwidth}}
   \caption{We show as a narrow dashed-dotted strip the predictions for the pion FF,
    obtained in the two-loop WI model, Eq.\ (\protect\ref{eq:Fpi.WI-Mod.NNLO}),
    using the minimal (left panel) and the improved (right panel) Gaussian models.
    The width of the strip is due to the variation of the Gegenbauer coefficients
    $a_2$ and $a_4$ (needed to calculate the collinear part
    $F_{\pi}^{\text{pQCD},(2)}(Q^2)$) in the corresponding shaded bands for the pion
    DA (indicated by the central solid line). Note that this dashed-dotted strip shows
    the effect of the $O(\mathcal A_2)$ correction only for the central solid curve
    of the shaded band. \protect\label{fig:FF.LD.2Loop}}
\end{figure}

It is interesting to note here,
that in the case of the one-loop approximation,
the relative error of the WI model (\ref{eq:Fpi-Mod.imp})
appears to be of the order of 10\%.
The relative weight of the $O(\alpha_s^2)$-contribution to the pion FF
is of the order of 10\%,
as has been shown in Refs.\ \refcite{BPSS04,AB08}.
Hence, the relative error of our estimate is of the order of
1\%---provided we take into account the $O(\alpha_s)$-correction
exactly via the specific choice of $s_0(Q^2)$,
as done in Eq.\ (\ref{eq:FF.LD.s0.App}).

The results obtained for the pion FF with our two-loop model,
i.e., Eq.\ (\ref{eq:Fpi.WI-Mod.NNLO}),
and using the effective LD thresholds
$s_{0}^\text{LD}(Q^2)$,
are displayed in Fig.~2.
We see from this figure
that the main effect of the NNLO correction
peaks at $Q^2\gtrsim4$~GeV$^2$, reaching the level of $3-10$\%.

\section{Conclusions}
We presented here the results for the spacelike pion form factor
obtained within the QCD SR approach,
using two different Gaussian NLC models.
These NLCs entail the decay of the nonperturbative OPE terms
at large $Q^2$.
These NLCs make the QCD SR stable and
enlarge the region
of its applicability
towards momenta as high as $10~\text{GeV}^2$.
The principal ingredients of our approach are,
besides the NLCs,
the $O(\alpha_s)$ spectral density, and
the analytic Shirkov--Solovtsov coupling
which is free of Landau singularities.
Our predictions for the pion FF
in the momentum range up to 10~GeV$^2$
are in a good agreement with the existing experimental data
of the Cornell\cite{FFPI-Cornell} and the JLab\cite{JLab08II} collaborations,
as well as with a recent lattice calculation.\cite{Brommel06}

We also showed here that the LD model for the pion FF
suffers from the threshold $s_0(Q^2)$ uncertainty.
We fixed this uncertainty by demanding that
the LD model should reproduce
the results of the Borel SRs with NLCs.\cite{BPS09}
Our results show that $s_0^\textbf{LD}(Q^2)$ grows with $Q^2$.

We also proved that the rough model for the matching function
in Ref.\ \refcite{BPSS04}
appears to be of a rather good quality ($\approx 10\%$).
We improved it here to reach the quality of $\approx 5\%$.
Using FAPT, and the improved matching function,
we estimated the NNLO correction to the pion FF
to be of the order of $\approx 3-10\%$.

\section*{Acknowledgments}
We are grateful to S.~V.~Mikhailov for helpful discussions.
Two of us (A.P.B. and A.V.P.) are indebted to Prof.\ Klaus Goeke
for the warm hospitality at Bochum University,
where part of this work was done.
The reported investigation was supported in part by the Deutsche
Forschungsgemeinschaft under contract DFG 436 RUS 113/881/0,
the Heisenberg--Landau Program, Grant 2009,
the Program ``Development of Scientific Potential in Higher Schools''
(projects 2.2.1.1/1483, 2.1.1/1539),
the Russian Foundation for Basic Research,
Grants No.\ ü~07-02-91557, 08-01-00686, and 09-02-01149,
and the BRFBR--JINR Cooperation Program, contract No.\ F08D-001.


\begin{thebibliography}{10}

\bibitem{ER80}
 A.~V. Efremov and A.~V. Radyushkin,
  \textit{Phys. Lett. B} \textbf{94},  245  (1980);
 \textit{Theor. Math. Phys.} \textbf{42},  97  (1980).

\bibitem{ER80tmf1}
 A.~V. Efremov and A.~V. Radyushkin,
 \textit{Theor. Math. Phys.} \textbf{44},  573  (1980).

\bibitem{Efr74yad}
 A.~V. Efremov,
 \textit{Yad. Fiz.} \textbf{19},  196  (1974).

\bibitem{EG74fph}
 A.~V. Efremov and I.~F. Ginzburg,
 \textit{Fortsch. Phys.} \textbf{22},  575  (1974).

\bibitem{LB79}
 G.~P. Lepage and S.~J. Brodsky,
 \textit{Phys. Lett. B} \textbf{87},  359  (1979).

\bibitem{CZ77}
 V.~L. Chernyak and A.~R. Zhitnitsky,
 \textit{JETP Lett.} \textbf{25},  510  (1977).

\bibitem{IL84}
 N. Isgur and C.~H. {Llewellyn Smith},
 \textit{Phys. Rev. Lett.} \textbf{52},  1080  (1984).

\bibitem{JK93}
 R. Jakob and P. Kroll,
 \textit{Phys. Lett. B} \textbf{315},  463  (1993).

\bibitem{BLM07}
 V. Braguta, W. Lucha, and D. Melikhov,
 \textit{Phys. Lett. B} \textbf{661},  354  (2008).

\bibitem{SSK}
 N.~G. Stefanis, W. Schroers, and H.-C. Kim,
 \textit{Phys. Lett. B} \textbf{449},  299  (1999);
 \textit{Eur. Phys. J. C} \textbf{18}, 137  (2000).

\bibitem{BPSS04}
 A.~P. Bakulev, K. Passek-Kumeri\v{c}ki, W. Schroers, and N.~G. Stefanis,
 \textit{Phys. Rev. D} \textbf{70},  033014, (2004);
 \textit{Phys. Rev. D} \textbf{70},  079906  (2004) Erratum.

\bibitem{NR82}
 V.~A. Nesterenko and A.~V. Radyushkin,
 \textit{Phys. Lett. B} \textbf{115},  410  (1982)

\bibitem{IS82}
 B.~L. Ioffe and A.~V. Smilga,
 \textit{Phys. Lett. B} \textbf{114},  353  (1982)

\bibitem{NR83}
 V.~A. Nesterenko and A.~V. Radyushkin,
 \textit{Phys. Lett. B} \textbf{128},  439  (1983).

\bibitem{BRS00}
 A.~P. Bakulev, A.~V. Radyushkin, and N.~G. Stefanis,
 \textit{Phys. Rev. D} \textbf{62},  113001  (2000).

\bibitem{MR-NLC}
 S.~V. Mikhailov and A.~V. Radyushkin,
 \textit{JETP Lett.} \textbf{43},  712 (1986);
 \textit{Sov. J. Nucl. Phys.} \textbf{49},  494  (1989);
 \textit{Phys. Rev. D} \textbf{45},  1754  (1992).

\bibitem{BR91}
 A.~P. Bakulev and A.~V. Radyushkin,
 \textit{Phys. Lett. B} \textbf{271},  223  (1991).

\bibitem{BMS01}
 A.~P. Bakulev, S.~V. Mikhailov, and N.~G. Stefanis,
 \textit{Phys. Lett. B} \textbf{508}, 279  (2001);
 \textit{Phys. Lett. B} \textbf{590}, 309  (2004)  Erratum.

\bibitem{BP06}
 A.~P. Bakulev and A.~V. Pimikov,
 \textit{Phys. Part. Nucl. Lett.} \textbf{4},  377  (2007).

\bibitem{BO04}
 V.~V. Braguta and A.~I. Onishchenko,
 \textit{Phys. Lett. B} \textbf{591},  267  (2004).

\bibitem{SS97-06}
 D.~V. Shirkov and I.~L. Solovtsov,
 \textit{Phys. Rev. Lett.} \textbf{79},  1209  (1997);
 \textit{Theor. Math. Phys.} \textbf{150},  132  (2007).

\bibitem{AB08}
 A.~P. Bakulev,
 \textit{Phys. Part. Nucl.} \textbf{40},  715  (2009).

\bibitem{Ste09}
 N.~G. Stefanis,
  arXiv:0902.4805 [hep-ph].

\bibitem{SVZ}
 M.~A. Shifman, A.~I. Vainshtein, and V.~I. Zakharov,
 \textit{Nucl. Phys. B} \textbf{147},  385  (1979).

\bibitem{MS93}
 S.~V. Mikhailov,
 \textit{Phys. Atom. Nucl.} \textbf{56},  650  (1993).

\bibitem{BPS09}
 A.~P. Bakulev, A.~V. Pimikov, and N.~G. Stefanis,
 \textit{Phys. Rev. D} \textbf{79},  093010  (2009).

\bibitem{DDM99}
 M. D'Elia, A. Di~Giacomo, and E. Meggiolaro,
 \textit{Phys. Rev. D} \textbf{59},  054503  (1999).

\bibitem{BM02}
 A.~P. Bakulev and S.~V. Mikhailov,
 \textit{Phys. Rev. D} \textbf{65},  114511  (2002).

\bibitem{DEM97}
 A.~E. Dorokhov, S.~V. Esaibegian, and S.~V. Mikhailov,
 \textit{Phys. Rev. D} \textbf{56},  4062  (1997).

\bibitem{PW96}
 M.~V. Polyakov and C. Weiss,
 \textit{Phys. Lett. B} \textbf{387},  841  (1996).


\bibitem{Brommel06}
 D. Brommel {\it et~al.},
  \textit{Eur. Phys. J. C} \textbf{51},  335  (2007).

\bibitem{FFPI-Cornell}
 C.~J. Bebek \textit{ et~al.},
 \textit{Phys. Rev. D} \textbf{9}, 1229 (1974);
 \textit{Phys. Rev. D} \textbf{13},   25 (1976);
 \textit{Phys. Rev. D} \textbf{17}, 1693 (1978).

\bibitem{JLab08II}
 G.~M. Huber {\it et~al.},
 \textit{Phys. Rev. C} \textbf{78},  045203  (2008)

\bibitem{Rad95}
 A.~V. Radyushkin,
 \textit{Acta Phys. Polon. B} \textbf{26},  2067  (1995).

\bibitem{IS83}
 B.~L. Ioffe and A.~V. Smilga,
 \textit{Nucl. Phys. B} \textbf{216},  373  (1983).

\bibitem{NR84}
 V.~A. Nesterenko and A.~V. Radyushkin,
 \textit{JETP Lett.} \textbf{39},  707  (1984).

\bibitem{BMS-APT}
 A.~P. Bakulev, S.~V. Mikhailov, and N.~G. Stefanis,
 \textit{Phys. Rev. D} \textbf{72}, 074014 (2005);
 \textit{Phys. Rev. D} \textbf{72}, 119908 (2005) Erratum;
 \textit{Phys. Rev. D} \textbf{75}, 056005 (2007);
 \textit{Phys. Rev. D} \textbf{77}, 079901 (2008) Erratum.

\bibitem{BKS05}
 A.~P. Bakulev, A.~I. Karanikas, and N.~G. Stefanis,
 \textit{Phys. Rev. D} \textbf{72}, 074015 (2005).

\end{thebibliography}

\end{document}